# HOM CHOICE STUDY WITH TEST ELECTRONICS FOR USE AS BEAM POSITION DIAGNOSTICS IN 3.9 GHZ ACCELERATING CAVITIES IN FLASH[§]

Nicoleta Baboi[#], Bastian Lorbeer, DESY, Hamburg, Germany
Pei Zhang, DESY, Hamburg, Germany; UMAN, Manchester, UK
Nathan Eddy, Brian Fellenz, Manfred Wendt, Fermilab, Batavia, USA

*Abstract*

Higher Order Modes (HOM) excited by the beam in the 3.9 GHz accelerating cavities in FLASH can be used for beam position diagnostics, as in a cavity beam position monitor. Previous studies of the modal choices within the complicated spectrum have revealed several options: cavity modes with strong coupling to the beam, and therefore with the potential for better position resolution, but which are propagating within all 4 cavities, and modes localized in the cavities or the beam pipes, which can give localized position information, but which provide worse resolution. For a better characterization of these options, a set of test electronics has been built, which can down-convert various frequencies between about 4 and 9 GHz to 70 MHz. The performance of various 20 MHz bands has been estimated. The best resolution of 20 μm was found for some propagating modes. Based on this study one band at ca. 5 GHz was chosen for high resolution position monitoring and a band at ca. 9 GHz for localized monitoring.

## INTRODUCTION

Higher Order Mode Beam Position Monitors (HOM-BPM) are devices that can be used to center the beam in accelerating cavities [1,2]. Since their principle relies on monitoring beam excited dipole modes, which are the main transverse component of the potentially damaging wakefields [3], they can help improve the quality of the charged particle beam. Moreover, they can be calibrated in terms of beam offsets, resembling a classical BPM.

HOM-BPMs have been built for the TESLA 1.3 GHz cavities at FLASH [1,4] and are routinely used for centering the beam. We are planning to build similar monitors for the 3.9 GHz cavities [5] in the same facility, often referred to as third harmonic cavities. The implementation is however much more complicated than in TESLA cavities for various reasons, briefly reviewed in the next section. Extensive studies have been made with the aim of defining the specifications of the HOMBPMs. The studies started with transmission measurements in each of the 4 cavities before and after installation in the cryo-module [6]. They continued with the dependencies of dipole modes on the transverse beam position, which had as result the identification of several regions in the HOM spectrum suitable for beam position monitoring [6,7]. The studies culminated with the examination of the potential of each region with a set of test electronics [8], which makes the subject of this paper. As a result, the specifications for the final electronics have been defined.

## FLASH

FLASH [4] is a Free Electron Laser facility generating short laser-like pulses with a wavelength between about 4 and 45 nm. It is also a test facility for the European X-ray FEL and the International Linear Collider. The first part of the linac is shown in Fig. 1. The photo-electric gun generates electron bunch trains with an energy of 5 MeV. These are accelerated by the first cryo-module, ACC1, containing 8 TESLA cavities. The subsequent cryo-module ACC39 is used to linearize the energy spread along the bunch generated by the non-linear accelerating field in ACC1 [9].

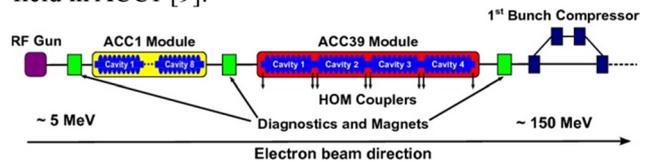

Figure 1: Schematics of the FLASH injector section.

## The Third Harmonic Cavities

Four 3rd harmonics cavities are installed in the ACC39 cryo-module (Fig. 2). They are denoted with C1 to C4, in the beam direction. There are 9-cells per cavity. Each cavity is equipped with an input coupler and 2 HOM couplers, placed in the connecting beam-pipes, to extract energy from the beam excited HOMs and thus reduce their effect on the beam quality.

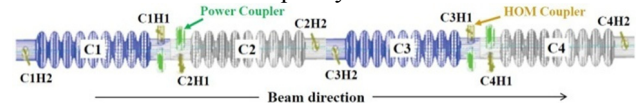

Figure 2: The four cavities in the ACC39 cryo-module. C2H1 means cavity 2, HOM-coupler 1.

The cavity design is inherited from the TESLA cavity. Figure 3 shows a picture of a 3.9 GHz cavity compared to a TESLA cavity. The main difference is the beam pipe diameter, which is larger than one third of the beam pipe in the TESLA cavity. This enables all HOMs to propagate along the module and therefore be damped by the couplers of all cavities.

___
[§]Work supported in part by the European Commission under the FP7 Research Infrastructures grant agreement No.227579
[#]nicoleta.baboi@desy.de

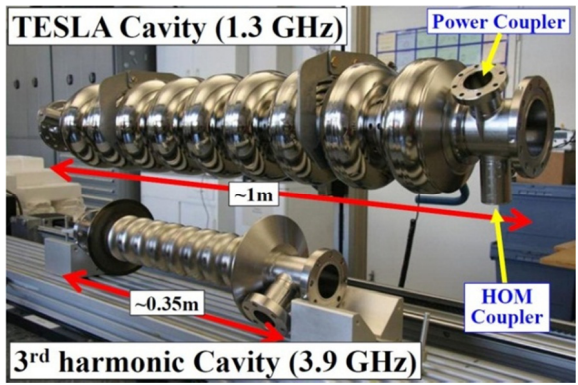

Figure 3: Picture of a 3.9 GHz cavity (down) and a 1.3 GHz cavity (up).

## The Principle of HOM-BPMs

Dipole modes are the main component of transverse HOMs and therefore have the highest damaging potential. Their strength depends linearly on the beam offset from the cavity center and therefore can be used for beam alignment and the monitoring of transverse beam positions, like classical cavity BPMs.

HOM-BPMs have been previously built for TESLA cavities at FLASH [1]. In that case, a dipole mode at ca. 1.7 GHz, one of the modes in the first dipole band, was used (see Fig. 4(a)). This mode has a high R/Q [10], a parameter which indicates the interaction strength of the mode with the beam. Thus a high R/Q has the potential for better position resolution.

Dipole modes in 3rd harmonic cavities are however overlapping with each other and are not easy to separate (see Fig. 4(b)). This is mainly due to the larger beam pipe, which makes the modes couple among cavities. This fact makes the beam position unable to be determined in each cavity, but only for the whole module. This is also true for the high R/Q modes.

Some modes in the fifth dipole band are found to be trapped inside each cavity [11,12] and could be used for local position diagnostics. However, these modes are located in an upper frequency range (ca. 9 GHz) and have only weak coupling to the beam (small R/Q values). The former requires careful electronics design and the latter will impact the position resolution of the HOM-BPM.

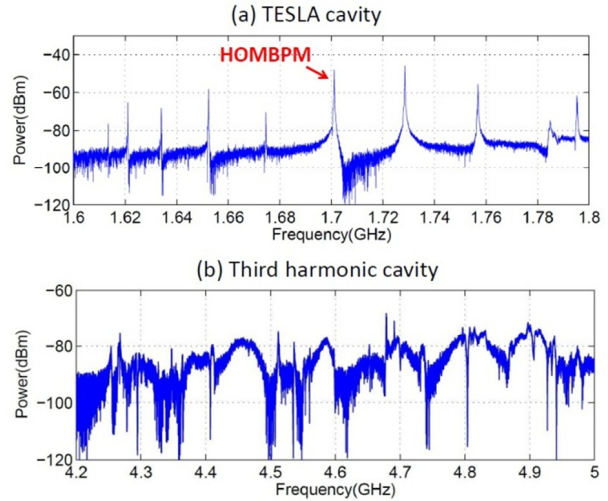

Figure 4: Beam-excited dipole modes in the 1st dipole band of a TESLA cavity and a 3rd harmonic cavity.

Since we cannot separate single modes, for the 3rd harmonic cavities multiple modes are used.. Initially this seemed to be a large disadvantage, but it may have some benefit. Using more modes means more information and has the potential for more resolution.

## MEASUREMENTS WITH THE TEST ELECTRONICS

### Results from Previous Studies

Along with extensive simulations [11-14], spectra measurements have been made with network and spectrum analyzers [6]. A beam-excited spectrum measured with a real-time spectrum analyzer (RSA) at coupler C1H1 is shown in Fig. 5 together with simulation results. The first two cavity dipole bands are between about 4.2 and 5.5 GHz, while the trapped cavity modes in the fifth dipole band are between 9.03 and 9.08 GHz.

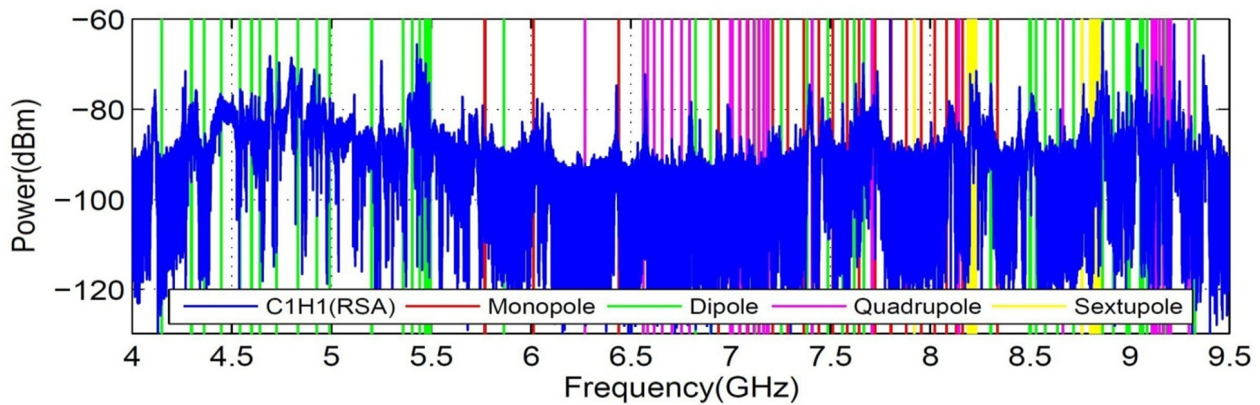

Figure 5: Beam-excited spectrum measured at coupler C1H1. Every 50 MHz from 4 GHz to 9.5 GHz was excited by a single electron bunch. The vertical lines indicate simulation results of an ideal cavity [11].

Studies on the behavior of the dipole modes on the beam position enabled finding of three regions of interest in the HOM spectrum promising for diagnostics, as summarized in Table 1 [2]. Beam-pipe dipole modes are localized between cavities and could deliver localized beam position estimations. However they have low R/Q's and are not cavity-based, therefore not relevant for beam alignment and the reduction of HOM effects through beam alignment. The first two cavity dipole bands contain modes with high R/Q's and thus have the potential for high resolutions, but they can only offer a measurement for the whole module. In the fifth dipole band there are trapped modes due to the geometry of end-cells and can provide a localized cavity-based beam measurement, but with low resolutions.

Table 1: Modal Options for HOM-BPM

| Mode type | Freq. (GHz) | Advantages | Disadvantages |
| --- | --- | --- | --- |
| Beam-pipe | ~ 4.1 | Localized | Low R/Q; Not cavity-based |
| Cavity/ 1st & 2nd band | 4.2-5.5 | High R/Q | Not localized |
| Cavity/ 5th band | 9.05-9.08 | Localized Cavity-based | Low R/Q |

Note that in our previous studies we could not study the resolution achievable with the various options, due to the limitations of the devices we used: oscilloscope and RSA. Therefore, in order to be able to make the final choice for the HOM-BPM electronics, flexible test electronics have been built.

*Setup*

The test electronics was designed to have the flexibility to study the various modal options of interest as well as accommodate the large mode bandwidths (BW) of these options. Its simplified block diagram is shown in Fig. 6. One of four different analog bandpass filters (BPF) can be connected into the chain to study the localized dipole beam-pipe modes at ca. 4.1 GHz, strong multi-cavity modes in the first dipole band at ca. 4.9 GHz and in the second dipole band at ca. 5.4 GHz, and the trapped cavity modes in the fifth dipole band at ca. 9 GHz. After filtering, the signal is mixed with a selectable local oscillator (LO) to an intermediate frequency (IF) of ca. 70 MHz. Then the IF signal is further filtered with a 20 MHz analog BPF to select the specific band of modes. In order to ensure that the possible remaining high frequency components of the IF signal generated during the mixing step are well suppressed, a lowpass filter (LPF) is applied to preserve only the frequencies below 105 MHz with a dominant component from 60 to 80 MHz. In Fig. 6, the "INPUT" is connected to the HOM signal through a multiplexer for easy switching between couplers. The 70 MHz IF out of the analog box is further split into two signals. One is digitized by a VME digitizer operating at 216 MS/s with 14 bit resolution along with a programmable FPGA for signal processing. The other IF signal is processed by a μTCA digitizer SIS8300 [15,16]. The VME digitizer is triggered by a 10 Hz FLASH beam trigger. Both the selectable LO and the digitizer clock are locked to the FLASH accelerator by using RF signals delivered from the master oscillator as a reference. This locking allows correct phase information of the digitized signal.

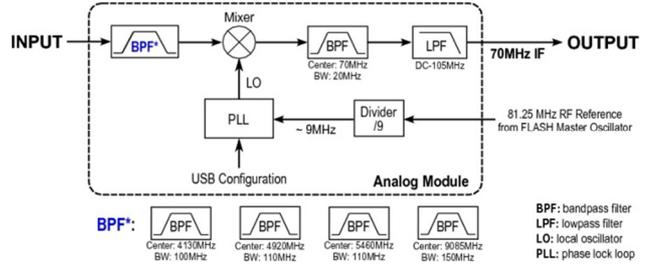

Figure 6: Schematics of the test electronics. One of the four BPFs was connected in front of the mixer during each measurement.

All the devices were set up outside the FLASH tunnel. Digitized data was collected from the VME digitizer with an EPICS software tool, while the beam charge, steerer current and BPM readouts were recorded synchronously from FLASH control system DOOCS. The data processing for position diagnostics is conducted offline using MATLAB.

This paper cites primarily the results from the VME digitizer. Results obtained with the μTCA digitizer can be found in Ref. [15].

An example of a signal measured with the VME digitizer for the input filter with a center frequency of 5.437 GHz and its Fourier transform are shown in Fig. 7.

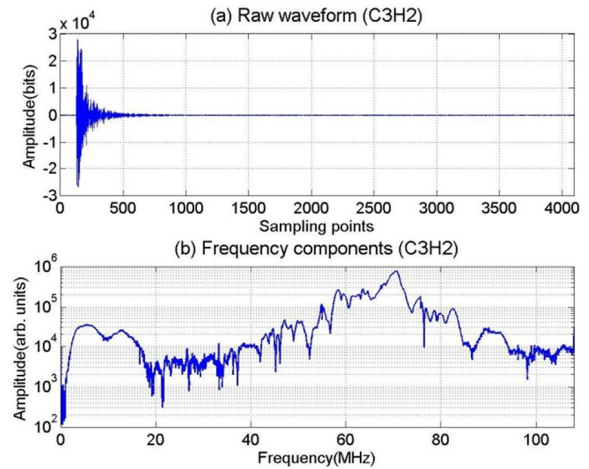

Figure 7: An example output signal (a) and its Fourier transform (b) measured from coupler C3H2 for a center frequency of 5.437 GHz.

## Measurements

The HOM frequency regions to be studied have been selected based on the previous measurements and simulation results as summarized in Table 2. The bandwidth was 20 MHz. A couple of tests with a bandwidth of 100 MHz have been made as well, with less promising results.

Table 2: Modal Options Selected from Previous Studies

| Mode type | Centre frequency (MHz) |
| --- | --- |
| Beam pipe | 4082, 4118 |
| 1st dipole band | 4859, 4904, 4940 |
| 2nd dipole band | 5437, 5464, 5482 |
| 5th dipole band | 9048, 9066 |

The schematic of the measurement setup is shown in Fig. 8. An electron bunch of approximately 0.5 nC was accelerated on-crest to ca. 150 MeV by ACC1 before entering the ACC39 module. Two steering magnets located upstream of ACC1 were used to produce horizontal and vertical offsets of the electron bunch in ACC39. Two beam position monitors (BPM-A and BPM-B) were used to record transverse beam positions before and after ACC39. By switching off the accelerating field in ACC39 and all quadrupoles nearby, a straight line trajectory of the electron bunch was produced between those two BPMs. Therefore, the transverse offset of the electron bunch in the module can be determined by interpolating the readouts of the two BPMs.

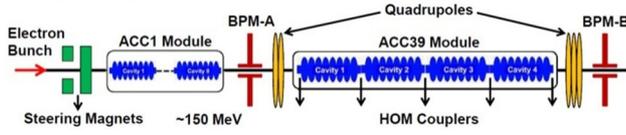

Figure 8: Schematic of measurement (not to scale; ACC1 is larger than ACC39).

In order to study the position dependencies of HOMs, we moved the beam in a 2D grid manner. Figure 9(a) shows the steerer current during the scan, while the readouts of the two BPMs are shown in Fig. 9(b) and Fig. 9(c). The samples in blue are used for calibration, and then validation samples were taken as shown in red to estimate the performance. The integrated HOM power over the frequency range shown in Fig. 7(b) is calculated for each beam position. Figure 9(d) shows the integrated power distribution measured from coupler C3H2. The position, which has minimum integrated power, is marked with white pentagon.

## Data Analysis

Singular Value Decomposition (SVD) has been applied to the matrix consisting of HOM waveforms in order to lower the system dimension by reducing the noise, and thus retain only the relevant information related to the beam position [7].

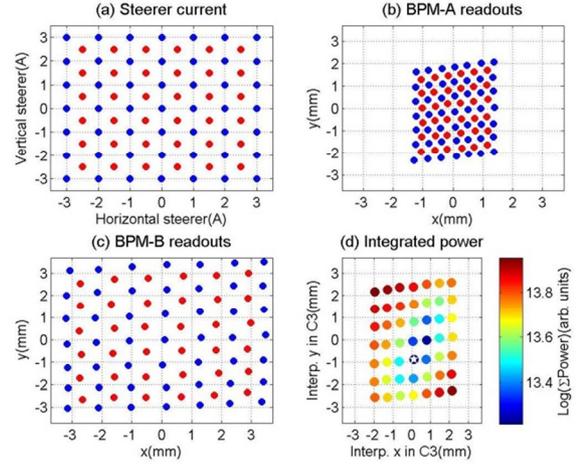

Figure 9: Steerer current (a) and readouts of BPM-A (b) and BPM-B (c) during a 2D grid-like beam scan (Calibration: blue; Validation: red). (d) Integrated power at different beam position. The center frequency was set to 5437 MHz with a BW of 20MHz.

The main SVD components have then be used to construct a "cleaner" waveform matrix, which has then been correlated to the matrix of beam positions where HOM signals have been measured. This means the center of the cavity for the localized modes, and the center of the cryo-module for the case of propagating modes. In this way the HOM signals are calibrated with beam offsets. By applying this calibration to newly measured HOM waveforms, a HOM-BPM is obtained. By comparing the reading of the HOM-BPM for the validation samples (red dots in Fig. 9) to the measurement interpolated from BPMs, an estimation of the measurement accuracy is made.

Figure 10(a) shows an example of the HOM-BPM readings (red) and the interpolated positions from BPMs (blue) when using the HOM signals shown in Fig. 7 for validation samples. The center frequency was 5.437 GHz, with a BW of 20 MHz. The differences between these two positions are shown in Fig. 10(b) as prediction errors:, 20 μm rms for $x$ and 22 μm rms for $y$.

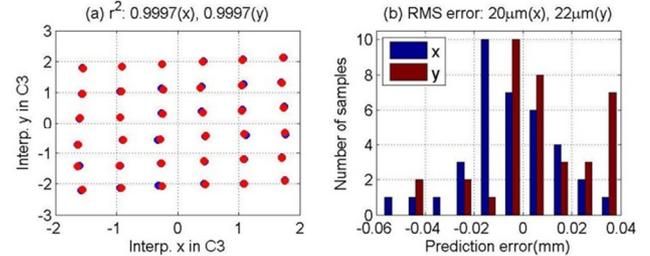

Figure 10: (a) Measurements (blue) and predictions (red) of the transverse beam position for a center frequency of 5.437 GHz. (b) Prediction errors.

It is worthwhile mentioning that this RMS error does not represent the usual resolution. We obtained smaller values when estimating this error for a smaller scan range of beam offsets. We expect to get even better results when measuring on beam jitter.

# RESULTS

## Localized Modes

### Beam pipe modes

For the beam pipe modes, the first filter in Fig. 6 was used. Only the µTCA-based digitizer was available at the time. The minimum number of SVD modes was chosen that leads to a smaller RMS error and it differs for the various couplers between 5 and 13. The RMS error obtained for the second frequency range, centered at 4.118 GHz, varies from coupler to coupler and is between 40 and 100 µm for the horizontal plane and between 80 and 180 µm for the vertical plane. The large error values, particularly for the y plane led to the elimination of this option early in the study.

### Trapped cavity modes

The fourth filter, centered at 9.085 GHz, was used to test the trapped modes. An ideal mathematical filter cut the propagating modes above 9.08 GHz. Similar results were obtained for the two frequency regions in the $5^{th}$ dipole band tested. RMS errors between 40 and 100 µm were obtained for all couplers and both transverse planes. 7 to 15 SVD modes were used. In spite of the large relatively large errors, we kept this option, since it can deliver localized, cavity based measurements. Also, we expect the resolution with the final electronics to be better.

## Propagating Cavity Modes

Surprisingly, the tested frequencies in the 1st dipole band performed slightly worse than the $2^{nd}$ band. The best results obtained were for a center frequency of 5.437 GHz. This may be due to the presence of high R/Q modes in this range, as predicted by simulations. The RMS errors are between 20 and 30 µm for the x plane and between 20 and 50 µm for the y plane (Fig. 11). Between 5 and 15 SVD modes were used. 10 - 20 µm rms is about the resolution of the BPMs used and therefore may be a limiting factor of our measurement. Due to the higher resolution, we chose to build an electronics version based on the second dipole band.

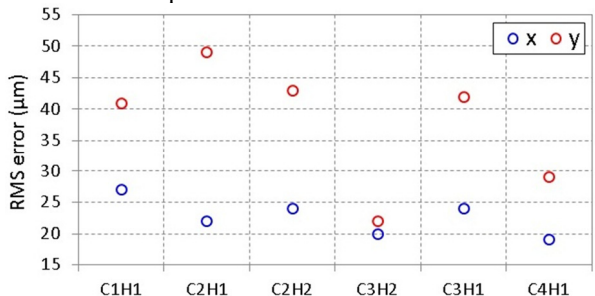

Figure 11: Position prediction accuracy for each coupler. The center frequency is 5437 MHz with a 20 MHz BW.

## Specifications for HOMBPM Electronics

Based on the test electronics results, the design of the final HOM-BPM electronics for the 3.9 GHz module has been finalized. Two couplers will be equipped with electronics centered at ca. 5440 MHz and a bandwidth of 100 MHz to provide high resolution positions in the module, and six couplers with electronics centered at ca. 9060 MHz, same bandwidth, to deliver local position inside each cavity. Since not all modes in the fifth dipole band are trapped [11,17], the frequency band has been carefully chosen in order to mitigate the contaminations of travelling modes.

# CONCLUSION AND OUTLOOK

Defining the frequency regions to be used for beam monitoring was no easy task, since all frequency regions in the spectra previously considered had advantages and disadvantages. Only with the flexible test electronics, which overcame the resolution limitations of the test devices previously used, oscilloscope and real-time spectrum analyzer, could these regions be further differentiated, and thus the specification of the final HOM-BPM electronics could be defined.

The HOM-BPM electronics is currently being designed by FNAL in collaboration with DESY. First tests should take place till spring 2013. The achieved resolution is expected to be at least as good as the one achieved with the test electronics.

Installing the electronics at FLASH is not the end of the project, since we experienced difficulties with the calibration stability. This may be due to phase drifts or even to drifts of the HOM frequencies. Further studies are required to understand this issue. Nevertheless as soon as the electronics is installed, it will already benefit beam alignment in ACC39.

# ACKNOWLEDGMENT

We are indebted to the accelerator operators and the FLASH team, as well as to the MCS group, for their support in preparing and conducting our measurements.